\documentclass[%
 reprint,
 amsmath,amssymb,
 aps,
 pre,
]{revtex4-2}

\usepackage{graphicx}

\usepackage{dcolumn}
\usepackage{bm}
\usepackage[colorlinks]{hyperref}

\begin{document}

\preprint{APS/123-QED}

\title{Permutation Entropy for the Characterization of the Attractive Hamiltonian Mean-Field Model}

\author{Melissa Fuentealba} \email{melifuentealba@udec.cl}
\author{Danilo M. Rivera} \email{danrivera@udec.cl}
\author{Roberto E. Navarro} \email{roberto.navarro@udec.cl}
\affiliation{
Departamento de Física, Facultad de Ciencias Físicas y Matemáticas, Universidad de Concepción, Concepción 4070386, Chile
}

\date{\today}

\begin{abstract}
  The Hamiltonian Mean-Field (HMF) model is a long-range interaction
  system that exhibits quasi-stationary states (QSS), which persist
  for long times before reaching thermodynamic equilibrium. These
  states are traditionally characterized by homogeneous/demagnetized
  or non-homogeneous/magnetized phase-space structures, separated by
  an out-of-equilibrium phase transition that depends on the initial
  energy $u_0$ and magnetization $M_0$ of the system. However, the
  magnetization also exhibits fluctuations around its mean value in
  time, which can provide additional insights into the nature of the
  QSS.  In this study, the permutation entropy $H$ and the statistical
  complexity $C$ are used as tools to characterize the dynamical
  properties of these magnetization fluctuations. It is found that
  most data points lie above the entropy-complexity curve for
  stochastic processes with a power-law spectrum (k-noise), suggesting
  that the magnetization retains more structure than purely stochastic
  processes. As the initial energy $u_0$ increases, both $H$ and $C$
  exhibit global minima that align closely with the critical energy
  $u^\ast$ separating magnetized and demagnetized QSSs. This agreement
  is particularly strong for $M_0 \lesssim 0.4$, where a first-order
  out-of-equilibrium phase transition has been reported for the HMF
  model. Below
  this transition, magnetized QSSs are associated with more ordered
  fluctuations, exhibiting fewer correlated structures.  Above this
  transition, demagnetized QSSs are characterized by fluctuations that
  shift toward more complex and more disordered dynamics, with an increasing
  number of correlated structures as $u_0$ increases.
\end{abstract}

\maketitle


\section{Introduction}

Systems with a large number $N$ of components interacting through
long-range forces can become trapped in a quasi-stationary state (QSS)
lasting for periods of times scaling as $t\sim N^{\alpha}$ before
reaching thermodynamic
equilibrium~\cite{Latora2001,Yamaguchi2004}. For example, while
studying the luminosity profiles of elliptical
galaxies,~\citet{Lynden-Bell1967} discovered that these systems were
in an apparent stationary state different from the equilibrium state
predicted by Chandrasekhar's theory~\cite{Chandrasekhar1960}, and
estimated relaxation times to reach true equilibrium would exceed the
current known age of the universe. This phenomenon is observed not
only in the dynamics of galaxies but also in various gravitational
systems~\cite{Dauxois2002b,Padmanabhan1990} and
plasmas~\cite{Levin2008}. In these systems, the dynamics are
collective, making the statistical treatment extremely complex to
apply due to the indeterminacy of the interaction distance.

The Hamiltonian Mean Field (HMF) model is one of the simplest models
that share characteristics with systems exhibiting long-range
interactions. It consists of $N$ spins that freely rotate in a plane
under the action of an infinite range cosine
potential~\cite{Pluchino2007}. The HMF model can also be viewed as an
approximation to a one-dimensional gravitational system, where the
two-particle interaction potential is replaced by the cosine of the
phase difference~\cite{Kumar2017,Miller2023}. This substitution
results in a phase-space of finite extent, making numerical
simulations more tractable. One of its main characteristics is that it
exhibits a non-collisional but violent relaxation process, leading to
out-of-equilibrium QSS~\cite{Levin2014}.

Typically, the QSSs of the HMF model have been studied by using a
waterbag (or step-like) initial distribution for the spin's
phase-space density, fully characterized by the total initial energy
$u_0$ of the system and its average magnetization $M_{0}$. In this
case, there exists an out-of-equilibrium phase transition separating
two types of QSS depending on $M_0$ and
$u_0$~\cite{Campa2009,Santini2022,Bachelard2008}: (A) An homogeneous
or demagnetized QSS, where the magnetization oscillates in time around
an average $M_\text{QSS}=0$, and traveling clusters in phase-space
formed probably by trapping of resonant
spins~\cite{Yamaguchi2011,Antoniazzi2007b}, and; (B) A non-homogeneous
or magnetized QSS, characterized by an average $M_\text{QSS}\neq0$ and
by a single non-traveling core-halo cluster in phase-space likely
formed from a Landau damping-like mechanism of density
waves~\cite{Pakter2011}.  Other out-of-equilibrium phase transitions
may exist, as more than two cluster in phase-space can be formed at
high enough values of $u_0$~\cite{Yamaguchi2011,Rivera2024}.

However, these QSS are not only characterized by the structures in the
spin's phase-space and the average magnetization.  Indeed, the
magnetization time series may present fluctuations with different
amplitude and modulation characteristics, depending on the initial
$u_0$ and $M_0$, that may contain information about the type of
QSS~\cite{Rivera2024}. Additionally, the HMF model is characterized by
chaotic behavior and anomalous diffusive behavior of its particles
over time~\cite{Latora1999, Yamaguchi2003, Pluchino2004a,
  Ginelli2011}. It also exhibits the formation of both ordered and
disordered spatial structures~\cite{Dauxois2000, Barré2002}, and the
clusters that emerge in phase-space can be considered universal
attractors~\cite{Pakter2011, Levin2014}. Moreover, the model
demonstrates self-organization~\cite{Barré2002, Bachelard2008} and
robustness in its QSSs~\cite{Jain2007,
  Pluchino2007, Staniscia2009}, while also being compatible with the
maximum entropy principle~\cite{Antoniazzi2007c,
  martelloni2016}. These features indicate that the HMF model shares
key characteristics with complex systems, particularly in its emergent
structures and non-trivial macroscopic behavior. While a deeper
investigation is needed to determine whether it exhibits traits such
as adaptability or self-tuning, its dynamics can nonetheless be
effectively analyzed using tools from complexity science.

In this work, we investigate whether the QSSs of the HMF model can be
characterized solely through magnetization fluctuations using the
permutation entropy~\cite{bandt2002}. Our goal is to determine if
these measures reveal meaningful information about the QSSs' nature
and their phase transitions. The permutation entropy is a widely used
measure to characterize ordered series, quantifying the balance
between order and disorder in a system, and assessing the dispersion
or randomness within a sequence.  It is computationally efficient,
robust against observational noise, and invariant to linear and
monotonic transformations~\cite{bandt2002}. A representation space
based on this measure is the complexity-entropy
plane~\cite{Rosso2007}, which allows distinguishing between chaotic
and stochastic systems. It provides insight into the temporal dynamics
of the system and helps identify whether the underlying process
follows a Gaussian or non-Gaussian distribution, as well as assess
different levels of correlation.

Several studies have successfully extended the use of the
complexity-entropy plane to a wide range of applications. For example,
plasma turbulence has been analyzed to differentiate chaotic behavior
in laboratory experiments from the stochastic nature of solar wind
fluctuations~\cite{Weck2015}, while financial time series have been
mapped onto this plane to distinguish between structured market
inefficiencies and random walk-like behavior~\cite{Zunino2010}. In
neuroscience, electroencephalogram recordings have revealed that
different brain states exhibit distinct entropy-complexity signatures,
with normal and pathological dynamics occupying separate regions of
the plane~\cite{Diaz2017eeg}. Furthermore, climate studies using
rainfall records and paleoclimate proxies have demonstrated the
method's ability to distinguish deterministic seasonal cycles from
random fluctuations~\cite{Silva2021}.  In space science, analyses of
geomagnetic indices have shown that auroral electrojet currents align
more closely with stochastic processes than with low-dimensional
chaos~\cite{Osmane2019}. These applications demonstrate the versatility of the
complexity-entropy plane in uncovering hidden patterns across
disciplines. Here, we apply this approach to analyze the
magnetization fluctuations in the HMF model.

\section{Numerical simulations of the Hamiltonian Mean-Field Model}
Here, we use the attractive HMF model, which describes the gyromotion
of $N$ fully coupled spins freely rotating in a plane as described by
the following Hamiltonian~\cite{Antoni1995,Campa2009}:
\begin{equation}
    \mathcal{H} = \frac{1}{2}\sum_{i=1}^{N}p_{i}^2 + \frac{1}{2N}\sum_{i,j=1}^N\left[1-\cos(\theta_{i}-\theta_{j})\right]\,, \label{eq:Hamiltonian}
\end{equation}
where $N$ is the total number of spins, $\theta_{i}$ is the angle of
rotation of the $i$-th spin, and $p_i$ its conjugate momentum. The
second term in Eq.~\eqref{eq:Hamiltonian} is independent of the
distance between spins, so it represents a potential of infinite
range. The Hamilton's equations for each spin are then:
\begin{align}
  \dot\theta_{i} &= p_{i} \,, &
  \dot p_{i}     &= -M_{x}\sin\theta_{i} + M_{y}\cos\theta_{i}\,, \label{eq:dotp}
\end{align}
where $M_x$ and $M_y$ are the components of the magnetization vector
$\vec{M}$, given by
\begin{align}
  M_{x} &= \frac{1}{N}\sum_{j=1}^{N} \cos\theta_j \,,
  &
    M_{y} &= \frac{1}{N}\sum_{j=1}^{N} \sin\theta_j \,.
            \label{eq:MxMy}
\end{align}

Note that the magnitude $M=\sqrt{M_x^2 + M_y^2}$ is a quantity
normalized to the interval $0\leq M \leq 1$. When $M=0$, the
orientation of all spins does not show a tendency towards a specific
angle, resulting in a homogeneous $\theta$-distribution. When $M=1$, the
spins are all oriented in the same direction.

To analyze the system dynamics, we perform numerical simulations based
on Eqs.~\eqref{eq:dotp} and~\eqref{eq:MxMy}. Our simulations involve a
large number of spins ($N=10^6$), and the simulation time $t$ is
discretized into $\mathcal{N}=2^{17}=131\,072$ iteration steps with a
temporal step $\Delta t=0.01$. The equations of motion~\eqref{eq:dotp}
are numerically solved through the classical second-order leap-frog
method, which ensures that the total energy is conserved over time,
with any possible oscillations remaining bounded.

The spins are initialized with random values of
$|\theta_i|\leq\theta_0$ and $|p_i|\leq p_0$ following a waterbag
distribution function, where $0\leq\theta_0\leq\pi$ and $p_0$ are
cut-offs in phase-space. By explicitly integrating
Eqs.~\eqref{eq:Hamiltonian} and~\eqref{eq:MxMy} over the phase-space
in the thermodynamic limit gives the initial total energy
$u_{0} = \frac{p_{0}^2}{6} + \frac{1}{2}(1-M_{0}^2)$ and the initial
magnetization $M_{x0}=M_0=\sin(\theta_{0})/\theta_{0}$ and
$M_{y0}=0$~\cite{Pakter2011}. Thus, the initial conditions are
characterized by either $\theta_0$ and $p_0$, or by $u_0$ and
$M_0$. Since $p_0^2\geq0$, states with $u_0<(1-M_0^2)/2$ are inaccessible. In the rest of this text, we use
$u_0$ and $M_0$ to characterize the initial conditions.

The equations of motion~\eqref{eq:dotp} are invariant under the
phase-space reflection $(\theta_i,p_i)\to
(-\theta_i,-p_i)$. Therefore, if the initial distribution function
satisfies $f(\theta,p,t=0)=f(-\theta,-p,t=0)$, this symmetry is
preserved for all $t\geq0$. This ensures that $M_y(t)=0$ at all times,
independently of the system size, and that the dynamics is fully
determined by $M_x(t)$. However, particle simulations introduce
numerical noise due to finite-size effects and numerical inaccuracies,
resulting in fluctuations in $M_y(t)$ that are nevertheless much smaller
in magnitude than the fluctuations in $M_x(t)$. Throughout this study, we
explicitly set $M_y(t)=0$, which allows us to reduce computational
costs without affecting the qualitative conclusions of the study. It
should be noted, however, that in contexts where the initial
distribution function does not satisfy the symmetry
$f(\theta,p,0)=f(-\theta,-p,0)$, it is necessary to calculate $M_y(t)$
to account for its role.

\begin{figure}[tb]
  \centering
  \includegraphics[width=0.482\textwidth]{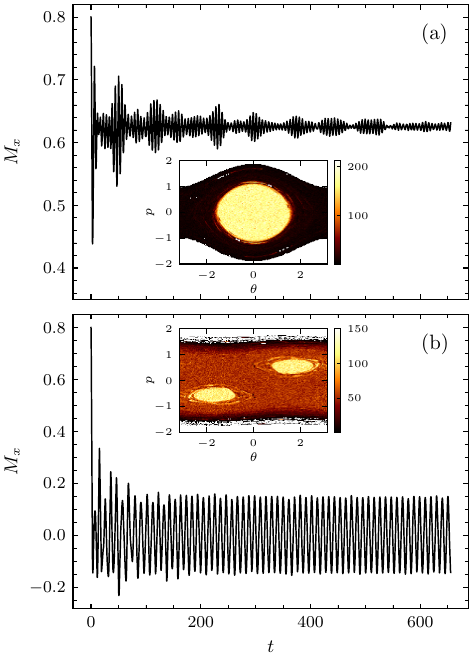}
  \caption{Temporal evolution of the magnetization $M_x(t)$ starting
    with $M_{0}=0.8$ for (a) $u_{0}=0.50$, and (b)
    $u_{0}=0.78$. Insets show the $\theta$-$p$ phase-space
    distributions during the QSS period at $t=650$, illustrating (a)
    the core-halo structure and (b) the nearly uniform distribution
    with traveling clusters. The color bar indicates the number of
    spins per bin in phase-space.  }
  \label{fig:magnetization}
\end{figure}

Figure~\ref{fig:magnetization} shows the time evolution of $M_x(t)$
for $M_0=0.8$ (or $\theta_0\simeq 1.131$) and the two cases $u_0=0.50$
($p_0\simeq1.39$) and $u_0=0.78$ ($p_0\simeq1.90$). Both cases quickly
relax within the first $20$ time steps from $M_x(0)=M_0$ to an average
magnetization of $M_\text{QSS}=0.63$ and $M_\text{QSS}=0$ in
Figs.~\ref{fig:magnetization}(a) and (b), respectively. This is a
violent relaxation of the magnetization as described
by~\citet{Lynden-Bell1967}. Subsequently, the system gets trapped in a QSS
where the magnetization $M_x(t)$ fluctuates around the average
$M_\text{QSS}$ values. In Fig.~\ref{fig:magnetization}(a), the
fluctuations in $M_x(t)$ appear to be modulated, while in
Fig.~\ref{fig:magnetization}(b), these fluctuations are uniform but
have higher intensity.

After $t=100$, the spin's phase-space density associated with
Fig.~\ref{fig:magnetization}(a) is characterized by a stationary,
single cluster-like structure, as shown in the insets. This structure
consists of a core limited to $|\theta|<2$ and $|p|<1.5$, surrounded
by a halo extending across the entire $\theta$ range, consistent with
observations by~\citet{Pakter2011}. Since this QSS structure is
bounded in phase-space, $M_\text{QSS}\neq0$. Additionally, smaller
structures orbit the core in a periodic clockwise motion, contributing
to the modulated fluctuations in $M_x$ seen in Fig.~\ref{fig:magnetization}(a).

In contrast, the inset of Fig.~\ref{fig:magnetization}(b) shows an
almost filled and nearly homogeneous phase-space, with two clusters
propagating in opposite directions. The upper cluster travels toward
increasing $\theta$, forming arms that spiral clockwise, whereas the
lower cluster evolves by following the phase-space symmetry
$f(\theta,p,t)=f(-\theta,-p,t)$. This configuration results in
wide-amplitude fluctuations in $M_x$ around $M_\text{QSS}=0$, as seen
in Fig.~\ref{fig:magnetization}(b).

\begin{figure}[tb]
  \centering
  \includegraphics[width=0.482\textwidth]{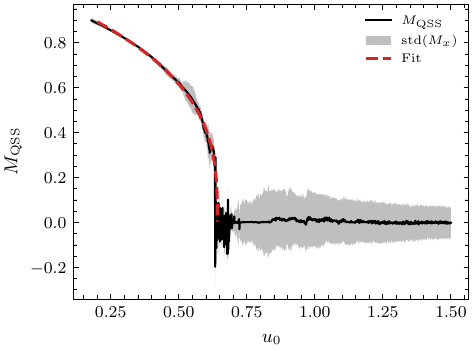}
  \caption{(black) Average magnetization $M_\text{QSS}$ during the
    quasi-stationary state as a function of the initial energy $u_{0}$
    for fixed $M_{0}=0.8$. (red) Power-law fit
    $M_\text{QSS} \sim (u^\ast-u_0)^\gamma$, where $u^\ast\approx0.64$
    is the critical energy separating states with $M_\text{QSS}>0$ and
    $M_\text{QSS}=0$, and $\gamma\simeq0.33$ is the critical
    exponent. (shaded gray) Standard deviation of the magnetization,
    indicating the amplitude fluctuations in $M_x$ around the average
    value $M_\text{QSS}$ during the QSS.  }
  \label{fig:average-M}
\end{figure}

To study only the window of time where the system is in a QSS, the
time series were truncated from $t=100$ (to remove the effects of the
relaxation process on $M_x$) to $t\approx655$. The black line in Fig.~\ref{fig:average-M} shows the average value
$M_\text{QSS}$, calculated as the mean of the truncated time series of
$M_x$ during the QSS, as a function of $u_0$ for a fixed value of
$M_0=0.8$. For low energies, $M_\text{QSS}>0$ is a decreasing function
of $u_0$, following a power-law dependency of the form
$M_\text{QSS} \sim (u^\ast - u_0)^\gamma$, where $\gamma = 0.33$ is
the critical exponent, and $u^\ast \approx 0.624$ is the critical
energy separating the magnetized ($M_\text{QSS}\neq0$) or inhomogeneous
QSS for $u_0<u^\ast$, and the nearly unmagnetized ($M_\text{QSS}=0$) or
homogeneous QSSs for $u_0>u^\ast$.

The shaded gray area in Fig.~\ref{fig:average-M} is the standard
deviation of $M_x$ as a function of $u_0$, representing the mean
amplitude of fluctuations around $M_\text{QSS}$. For $u_0 < u^\ast$,
the amplitude fluctuations are significantly lower compared to
$u_0 > u^\ast$, which is consistent with expectations for systems with
higher energy or temperature generally exhibiting larger
fluctuations. In the region $u^\ast < u_0 < 0.7$, the mean
$M_\text{QSS}$ is not exactly zero, and the amplitude fluctuations
remain relatively low. Beyond $u_0 > 0.7$, the amplitude of
fluctuations grows with $u_0$, reaching a maximum around
$u_0 \approx 0.85$, and then decreases slightly and monotonically for
larger values of $u_0$.

These observations suggest a strong correlation between the amplitude
of fluctuations and the QSS across the out-of-equilibrium phase
transition at $u^\ast = 0.64$. Below the critical energy, the
amplitude of fluctuations are much smaller compared to those above the
critical energy, reinforcing the idea that the transition from
magnetized to unmagnetized QSS is accompanied by a qualitative shift
in the dynamical properties of $M_x(t)$. In this work, we study the
time series of the magnetization $M_x$ in the QSS for different
initial conditions $u_0$ and $M_0$, without relying on phase-space
information. While several studies have analyzed the microscopic
phase-space dynamics of the HMF model, e.g.~\citet{Bachelard2008}
and~\citet{Benetti2012}, our goal is to characterize the QSS through
the fluctuations in $M_x(t)$.

This approach is particularly relevant in practical scenarios where
access to full phase-space data is limited. For instance,
quasi-thermal noise spectroscopy is a robust method for extracting
electron plasma parameters from electrostatic fluctuations, enabling
in-situ plasma measurements without active
probes~\cite{MeyerVernet2017} or where particle detectors are
ineffective due to low-energy electrons~\cite{MeyerVernet1986}. More
generally, the fluctuation-dissipation theorem formalizes how
equilibrium noise, such as Johnson-Nyquist voltage
fluctuations~\cite{qu2019} or density variations in turbulent plasmas,
relates directly to thermodynamic parameters like temperature,
compressibility, and transport coefficients. However, the
fluctuation-dissipation theorem applies strictly to systems in
thermodynamic equilibrium or close to it. For non-equilibrium QSSs,
complexity measures such as permutation entropy~\cite{bandt2002},
statistical complexity~\cite{lopezruiz1995}, and other statistical
methods provide a more suitable framework for detecting chaotic
dynamics and hidden correlations.

\section{Permutation Entropy and Statistical Complexity}
We briefly describe how to evaluate the permutation
entropy~\cite{bandt2002} for the fluctuations in the
magnetization. Consider the time series $M_i = M(t=i\Delta t)$ with
$0\leq i\leq \mathcal{N}$ an integer representing the iteration time
step. We subdivide this series into partitions of $d$ elements as
\begin{equation}
  W_{p} = \{ M_{p},\,
  M_{p+\tau},\,
  M_{p+2\tau}, \dots,\,
  M_{p+(d-1)\tau} \}\,,
\end{equation}
where $1\leq p\leq n$ is the partition index,
$n=\mathcal{N} - (d-1)\tau$ is the total number of possibly
overlapping time windows, $\tau\geq 1$ is the embedding delay denoting
a temporal shift between windows, and $d>1$ is the embedding window
size.

Next, we construct $n$ permutation sequences
$\pi_{p}=\{r_0,\,r_1,\,\dots,\,r_{d-1}\}$ of the indices
$0\leq r_i\leq d-1$ that would sort the elements of each partition
$W_p$ in ascending order. Then, we calculate the relative frequency in
which any given sequence $\pi$ is found among all the permutation
sequences $\{\pi_p\}_{p=1\dots{n}}$, i.e.
\begin{equation}
\rho(\pi)=\frac{1}{n} \sum_{p=1}^n \delta(\pi, \pi_p) \,,
\end{equation}
where $\delta(\pi, \pi_p)$ is a delta-like function equal to $1$ if
$\pi=\pi_p$, and $0$ otherwise.

Finally, the permutation entropy is the Shannon entropy of the
probability distribution $\rho(\pi)$, which measures uncertainty or
information content characterizing the probability distribution of
ordinal patterns in a time series. Here, we use the normalized
permutation entropy as~\cite{Pessa2021}
\begin{equation}
  \label{eq:entropy}
  H(\rho) = -\frac{1}{\log_2(d!)} \sum_\pi \rho(\pi) \log_2 \rho(\pi) \,,
\end{equation}
so that $0\leq H \leq 1$, where $H\simeq0$ suggests that there exists
a predominant pattern, indicating more predictable data or a somewhat
regular dynamics (e.g. monochromatic series); and $H\simeq1$ indicating a greater uncertainty or
disorder in the time series (e.g. white noise).

On the other hand, the complexity-entropy plane was used by
\citet{Rosso2007} as a framework to distinguish between chaotic and stochastic time
series. It is based on the statistical complexity measure proposed by
\citet{lopezruiz1995}, given by
\begin{equation}
  \label{eq:complexity}
  C(\rho) = \frac{D(\rho,U)}{D_\text{max}} H(\rho)\,,
\end{equation}
where $U=1/d!$ is the uniform distribution for all possible sequences
$\pi$, $D_\text{max}$ is a normalization constant, and $D(\rho,U)$ is
the extensive Jensen-Shannon divergence \cite{Martin2006} given by
\begin{equation}
  \label{eq:Jensen-Shannon}
  D(\rho,U) = S\left(\frac{\rho+U}{2}\right) - \frac{1}{2} S(\rho) - \frac{1}{2} S(U)\,,
\end{equation}
where $S(\rho)=-\sum_\pi \rho(\pi) \log_2 \rho(\pi)$ is the
unnormalized Shannon entropy.

The statistical complexity Eq.~\eqref{eq:complexity} is related to the
interaction between the amount of information a system possesses and
its imbalance, quantifying the degree of correlational structure in
the series. For the numerical analysis of the permutation entropy and
the statistical complexity, the open-source \textit{ordpy}
module~\cite{Pessa2021} of the \textit{Python} programming language
was used.

\begin{figure}[tb]
  \centering
  \includegraphics[width=0.482\textwidth]{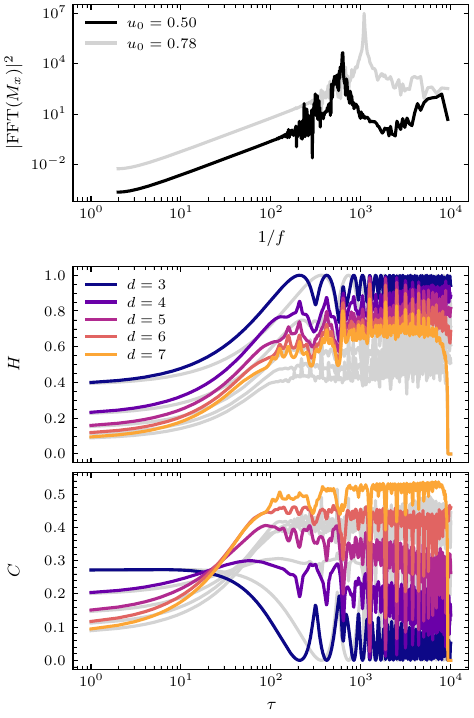}
  \caption{(top) Log-log plot of the power spectrum of $M_x(t)$ as a
    function of the inverse of the frequency $1/f$, corresponding to
    characteristic oscillation timescales, for $M_0=0.8$ and the cases
    shown in Fig.~\ref{fig:magnetization}. (middle) Semi-log plot of
    the permutation entropy $H$ and (bottom) the statistical complexity
    $C$ of $M_x(t)$ during the QSS as a function of the embedding
    delay $\tau$. The colored lines represent different embedding
    dimensions $d$ for the case $u_0=0.50$, as in
    Fig.~\ref{fig:magnetization}(a). The gray lines show results for
    the case $u_0=0.78$, as in Fig.~\ref{fig:magnetization}(b).}
  \label{fig:fft}
\end{figure}

Figure~\ref{fig:fft} shows an analysis of the entropy and complexity
calculations for the cases shown in Fig.~\ref{fig:magnetization}, as
functions of the embedding delay $\tau$ and the embedding dimension
$3\leq d\leq 7$. The choice of the embedding delay $\tau$ determines
how successive values in the time series are spaced when constructing
ordinal patterns. If $\tau$ is too small, successive values may be
nearly identical, reducing the ability to detect meaningful
structures. On the other hand, excessively large values of $\tau$ may
overlook relevant short-term correlations and distort the underlying
dynamics.  A reasonable approach is to set $\tau$ based on the
system's intrinsic time scales, which can be inferred from its
temporal correlations or characteristic oscillations. The top panel of
Fig.~\ref{fig:fft} shows the power spectrum of $M_x(t)$ as a function
of the inverse of the frequency $1/f$, which represents characteristic
oscillation timescales. The spectrum for $u_0 = 0.50$ displays a
double-peaked structure with prominent features at $1/f \sim 600$ and
$1/f \sim 8000$ time steps, as expected from the modulated oscillatory
signal shown in Fig.~\ref{fig:magnetization}(a).  In contrast, the
spectrum for $u_0 = 0.78$ exhibits a single dominant peak at
$1/f \sim 1000$ time steps.

The middle panel of Fig.~\ref{fig:fft} displays the permutation
entropy $H$, Eq.~\eqref{eq:entropy}, as a function of the embedding
delay $\tau$. The colored lines represent different embedding
dimensions $d$ focused on the case $u_0=0.50$, while the gray lines
are the case $u_0=0.78$. For all values of $d$, the entropy increases
approximately as $H\propto(\log_{10}\tau)^3$ at small $\tau$. This growth saturates
around $\tau \sim 200$, before the first characteristic timescale
($1/f \sim 600$). This behavior suggests that at short timescales
(relative to the dominant oscillation period), the ordinal patterns
become gradually more decorrelated as $\tau$ increases. Beyond
$\tau\gtrsim200$, $H$ stabilizes near its maximum possible value,
except for some variations. Thus, for timescales larger than the
dominant oscillation period, the patterns are fully decorrelated, and
increasing $\tau$ further does not introduce additional structure
complexity. The entropy systematically decreases with $d$, but its
overall trend as a function of $\tau$ remains consistent across
different embedding dimensions.

The bottom panel of Fig.~\ref{fig:fft} shows the statistical
complexity $C$, Eq.~\eqref{eq:complexity} as a function of $\tau$. A
notable distinction from the entropy behavior is that for $d = 3$, the
complexity decreases with $\tau$, whereas for $d \geq 4$, $C$
increases until saturating around $\tau \sim 200$. This suggests that
a low embedding dimension like $d=3$ might not capture enough
complexity in the time series, due to the limited number of possible
patterns, resulting in an underestimation of the system's dynamical
structure. This can artificially inflate the entropy $H$ while
reducing the complexity $C$, since patterns may appear more random
than they actually are. After $\tau\gtrsim200$, $C$ slowly decreases
with $\tau$, indicating that the structural richness of ordinal
patterns does not change significantly across different embedding
delays, and high values of $\tau$ do not introduce new correlations.

Notice that the case $d=6$ is nearly identical to $d=7$ for both $H$
and $C$. Although higher values ($d \geq 7$) allow for more detailed
ordinal patterns, they require significantly longer time series
($\mathcal{N} \gg d!$) to ensure proper pattern
sampling~\cite{Cuesta-Frau2019}.  Since increasing $d$ beyond 6 does
not significantly change the results, we select $d = 6$ as the optimal
balance between capturing the complexity of $M_x(t)$ and computational
feasibility.

\begin{figure*}[tb]
  \centering
  \includegraphics[width=\textwidth]{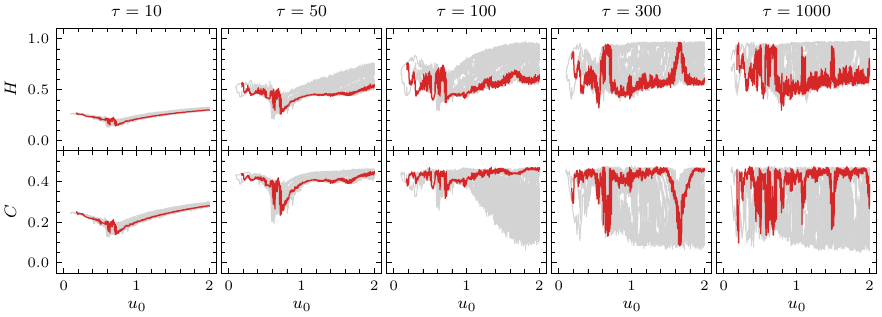}
  \caption{(top) Permutation entropy $H$ and (bottom) statistical
    complexity $C$ of the magnetization time series $M_x(t)$ during
    the QSS, as functions of the initial energy $u_{0}$. Each column
    corresponds to a different embedding delay $\tau$ with embedding
    window $d=6$. The gray lines show results for different initial
    magnetization values $M_0$, while the red line highlights the case
    $M_0 = 0.8$.}
  \label{fig:entropy}
\end{figure*}
To systematically study the effect of $\tau$, we analyzed the
permutation entropy Eq.~\eqref{eq:entropy} and the statistical
complexity Eq.~\eqref{eq:complexity} of the magnetization $M_x(t)$ for
different initial conditions and for $d=6$.  Figure~\ref{fig:entropy}
shows the results for eleven equally spaced values of the initial
magnetization, $M_0=\{0.0,\, 0.1,\, 0.2,\, \dots,\, 1.0\}$. For each
value of $M_0$, we ran over $1\,000$ simulations with initial energies in
the range $u_\text{min}<u_0<2.0$, where $u_\text{min}=(1-M_0^2)/2$ is
the threshold of inaccessible states. The red line highlights the case
$M_0=0.8$, while the gray lines correspond to the other cases. A
Fourier analysis of $M_x(t)$ revealed characteristic periods
predominantly between $200$ and $1000$ time steps in these
simulations. Based on this, we evaluated the permutation entropy across
various values of $1\leq\tau\leq1\,000$. Figure~\ref{fig:entropy} is
thus organized into five columns, each depicting a representative case
from the full set of $\tau$ values analyzed.

The general trend observed is that both $H$ and $C$ tend to increase
as $\tau$ increases, which is coherent with the description of Fig.~\ref{fig:fft}(middle). For very small embedding delays, e.g. $\tau=1$
not shown here, both $H$ and $C$ remain almost constant with respect to $u_0$, with their
trends closely resembling those for $\tau = 10$. For cases with
$\tau \leq 10$, $H$ and $C$ appear relatively insensitive to the
initial magnetization $M_0$, as indicated by the narrow spread of the
gray curves in the figure. Their variations are primarily determined
by the energy $u_0$. As $\tau$ increases beyond $10$, the dependence
on $M_0$ becomes more pronounced, with the gray curves spreading
further apart. The entropy $H$ tends to be highest for $M_0 = 0.0$ in
all cases, and it gradually decreases for larger $M_0$, reaching its
lowest values around $M_0=0.8$. A similar trend is observed for the
complexity $C$. For small $\tau\leq100$, $C$ tends to be higher for lower
$M_0$, whereas for large $\tau>100$, the opposite occurs.

Examining the dependence on $u_0$, we observe that $H$ exhibits a
decreasing trend with $u_0$ for $1 \leq \tau \leq 200$, except for
some erratic variations, and reaches a noticeable dip around
$u_0 \approx 0.7$. Beyond $u_0\gtrsim0.7$, $H$ exhibits a nearly smooth increasing trend. When $\tau>200$, this dip remains visible but $H$
develops stronger variations across $u_0$. The position of the minimum
entropy remains near $u_0 \approx 0.7$ for $M_0 = 0.8$ for all $\tau$
values. However, this minimum shifts to lower $u_0$ as $M_0$
decreases. Notice that the dip occurs near the critical energy
$u^\ast \approx 0.64$ that separates the magnetized and demagnetized
QSS as shown in Fig.~\ref{fig:average-M}.

The complexity $C$ follows a similar functional dependence on $u_0$ as
$H$, especially for small $\tau$. However, for larger $\tau \geq 100$,
$C$ approaches an upper bound around $C \approx 0.45$, except for
localized dips at certain values of $u_0$. The variations in $C$ also
increase with $\tau$.

The variations at high $\tau$ could indicate a breakdown of ordinal
pattern recognition due to excessive separation between embedded
points in the time series. When $\tau$ is too large, correlations
between time-delayed values weaken, leading to irregularities in the
entropy and complexity measures.  This reinforces the need to balance
$\tau$ when analyzing ordinal patterns. Finally, the relatively smooth
behavior of $H$ and $C$ for low $\tau$ ($\tau = 1$ to $200$) suggests
that the permutation entropy is capturing some structure at timescales
below the dominant period of the magnetization fluctuations. While the
exact nature of this structure requires further investigation, it may
relate to short-term correlations or subharmonic oscillations within
the quasi-stationary states of the system.

It is worth mentioning that, since the permutation entropy $H$ only
concerns patterns in the ordering of $M_x$, the entropy of $M_x$
and its fluctuations $\delta M_x = M_x - M_\text{QSS}$ are identical, i.e.  $H(M_x)=H(\delta M_x)$. This implies that $H$,
and by extension $C$, is independent of the amplitude scaling of the
fluctuations and does not capture information about the mean value
$M_\text{QSS}$. However, $H$ can still reflect spectral properties of
$\delta M_x$. Indeed, for time series dominated by periodic or
quasi-periodic signals, $H$ tends to be lower due to the regular
repetition of patterns. Conversely, for random time series such as white
noise, $H$ is higher, as the ordinal patterns become uniformly
distributed.

\begin{figure*}[tb]
\centering
\includegraphics[width=\textwidth]{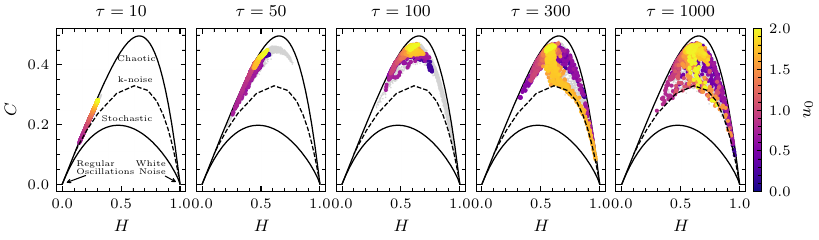}
\caption{Complexity-entropy plane for the magnetization time series
  $M_x(t)$ during the QSS regime. Each column corresponds to a
  different embedding delay $\tau$ and embedding window $d=6$. The
  color bar indicates the initial energy $u_0$ for the case $M_0=0.8$,
  corresponding to the red lines in Fig.~\ref{fig:entropy}, while the
  gray points represent data for other values of $M_0$. The solid
  upper and lower curves represent the theoretical maximum
  $C_\text{max}$ and minimum $C_\text{min}$ bounds of the statistical
  complexity~\cite{Martin2006}. The dashed curve represents the
  entropy-complexity line for stochastic processes with a power-law
  spectrum $1/f^k$ (k-noise), ranging from white noise ($k = 0$) to
  strongly correlated noise ($k = 6$), calculated using $\tau = 1$ in
  all panels.}
\label{fig:plane}
\end{figure*}

\section{Complexity-Entropy Plane}
To further analyze the dynamical properties of $M_x(t)$, we present in
Fig.~\ref{fig:plane} the complexity-entropy plane for different
embedding delays $\tau$. This representation allows us to distinguish
between stochastic and chaotic-like systems~\cite{Rosso2007}.
Figure~\ref{fig:plane} highlights the complexity-entropy plane for
simulations with $M_{0}=0.8$, corresponding to the red lines in
Fig.~\ref{fig:entropy}. Each colored point represents a pair of
entropy $H$ and complexity $C$ values computed from $M_x(t)$ in the
QSS for a given $u_0$, indicated by the colors in the side bars.
Other values of $M_0$ are also included as gray points.

The black lines represent the theoretical minimum and maximum
statistical complexity $C$ for a given entropy $H$ at
$d=6$~\cite{Martin2006}. The minimum complexity curve $C_\text{min}$ corresponds to
systems with the least structural correlation for a given entropy. The
maximum complexity curve $C_\text{max}$, on the other hand, corresponds to systems
with the highest possible structural richness, often associated with
chaotic or intermittent dynamics. For comparison purposes, we include a dashed line representing the
entropy-complexity relation for stochastic processes with a power-law
spectrum $1/f^k$ (k-noise), ranging from white noise ($k = 0$) to
strongly correlated noise ($k = 6$). 

For $\tau = 10$, the data points in Fig.~\ref{fig:entropy} cluster in
the lower-left region of the complexity-entropy plane near $0.1<H<0.3$
and $0.1 < C < 0.3$, following an approximately linear dependence $C\simeq
H$. Most points lie above the k-noise curve, very close to
$C_{\text{max}}$, indicating that the system retains a high degree of
correlated structures at short timescales. The gray and colored points
are closely grouped, suggesting that the initial magnetization $M_0$
has little influence on $H$ and $C$ for small $\tau$. Although not
show here, the points are even more clustered for $\tau=1$ to $5$, with
$C\approx H\approx 0.1$.

For $\tau = 50$ and $\tau=100$, the data points shift toward higher
entropy and complexity values, moving rightward and upward in the
plane. The spread in entropy and complexity increases, and the
distinction between different energy levels becomes more visible. The
trend where $H$ and $C$ decrease up to $u_0 = 0.7$ and then increase
is now clearly observed. For $u_0 < 0.7$, points shift toward lower
entropy and complexity, whereas for $u_0 > 0.7$, points move back
toward higher entropy and complexity, along the $C_{\text{max}}$
curve. Although the highlighted case $M_0=0.8$ exhibit points bounded
by $0.3<H<0.8$ in the plane, the gray points spread further, reaching
entropies as high as $H = 0.8$ for $\tau=50$, and $H=1$ for $\tau=100$,
but still remain in the upper quarter of the plane.

For $\tau = 300$ and $\tau=1000$, most of the upper half of the
complexity-entropy plane is now occupied by the colored points. While
the previous well-defined structure with a single turning point at
$u_0 = 0.7$ is no longer dominant, low-energy states still cluster in
the upper quarter near $C_{\text{max}}$, and high-energy states tend
to cluster near the k-noise curve. At very high energies
($u_0 \approx 2$), the points concentrate around $H \approx 0.5$ and
$C \approx 0.45$, forming a distinct grouping. For $\tau = 1000$, the
data points now spread across the entire upper half of the plane,
occupying all regions between the k-noise curve and
$C_{\text{max}}$. The boundaries between different energy regimes
become less distinct, and the dispersion in entropy and complexity is
at its maximum.

Across all values of $\tau$, most data points remain above the k-noise
curve, indicating that the magnetization time series retains more
structure than purely stochastic processes. However, while many points
are close to $C_{\text{max}}$, which is often associated with chaotic
or intermittent dynamics, the dispersion of points at large $\tau$
suggests that the system is not purely chaotic in a deterministic
sense. Instead, the spreading in entropy and complexity at large $\tau$
likely results from decorrelation effects, where distant points in the
time series become weakly correlated, leading to an apparent increase in
disorder. This suggests that while fluctuations in $M_x(t)$ maintain
structured patterns, the system's behavior at large timescales may be
closer to correlated stochastic processes rather than purely
deterministic chaos.

\begin{figure}[tb]
  \centering
  \includegraphics[width=0.482\textwidth]{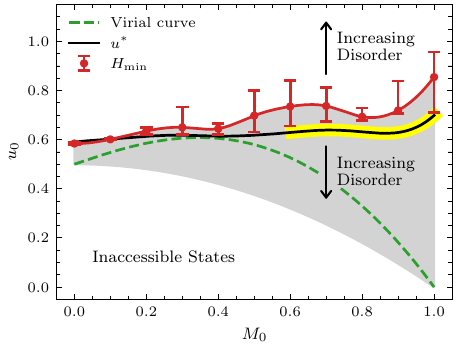}
  \caption{(red line) Initial energies $u_{H\text{min}}$ for
    which the permutation entropy $H$ in Fig.~\ref{fig:entropy} is
    minimized, as a function of $M_0$. The QSS exhibits the highest
    degree of ordered patterns in $M_x(t)$ near this line. As $u_0$
    moves away from it, $H$ increases and $M_x(t)$ becomes
    progressively more disordered. The vertical bars indicate the
    dispersion of energies for a given $M_0$ due to calculation at
    different embedding delays $\tau$. (black line) Line of
    first-order phase transitions, as reported by~\citet{Benetti2012},
    corresponding to the critical energies $u^\ast$ extracted from
    Fig.~\ref{fig:average-M}. The QSS is magnetized
    ($M_\text{QSS} > 0$) below this line, and demagnetized
    ($M_\text{QSS} = 0$) above it. (Yellow line) Points where the
    order of the phase-transition becomes unclear, according
    to~\citet{Benetti2012}.  (green dashed line) The virial curve of
    the system, also reported by~\citet{Benetti2012}. The lower white
    area represents the states $u_0<\frac{1}{2}(1-M_0^2)$ that are
    physically inaccessible in the HMF simulations.}
  \label{fig:transition}
\end{figure}

Lastly, we investigate how the energy $u_0$ at which the permutation
entropy $H$ reaches its minimum depends on
$M_0$. Figure~\ref{fig:transition} provides a comprehensive overview
of the initial energy $u_0$ versus the initial magnetization $M_0$,
highlighting key features of the QSS in the HMF model. In particular,
the red line represents the median energies
$u_{H\text{min}}=\langle u_{H\text{min}}(\tau)\rangle$ for which the
permutation entropy $H$ reaches its global minimum for a given $M_0$.
Since $H_\text{min}$ varies with the embedding delay $\tau$, as shown
in Fig.~\ref{fig:entropy}, each value of $M_0$ is associated with
multiple values of $u_{H\text{min}}(\tau)$. To extract a
representative transition energy, we compute the median of
$u_{H\text{min}}(\tau)$ across all values of
$1 \leq \tau \leq 1\,000$. The 16th and 84th percentiles of
$u_{H\text{min}}(\tau)$ are then used to estimate its variability due
to the dependence on $\tau$, resulting in the uneven error bars drawn
in Fig.~\ref{fig:transition}. These percentiles correspond to one
standard deviation from the mean if $u_{H\text{min}}(\tau)$ follows a
normal distribution over $\tau$.

Near the red line in Fig.~\ref{fig:transition}, the magnetization
fluctuations $M_x(t)$ exhibit the most structured evolution, with the
lowest entropy values indicating reduced randomness and stronger
temporal correlations. As $u_0$ deviates away from the red line in
either direction, $H$ increases, reflecting a progressive loss of
order and structure in $M_x(t)$.

The black line in Fig.~\ref{fig:transition} represents the critical
energy $u^\ast$, where the system transitions from magnetized to
demagnetized QSSs, as extracted from Fig.~\ref{fig:average-M}. A
striking feature is that the red and black lines almost coincide,
despite being deduced from different methodologies. The $H_\text{min}$
red line aligns closely with the $u^\ast$ black line for
$M_0 \lesssim 0.4$ where a first-order out-of-equilibrium phase
transition has been identified for the HMF
model~\cite{Benetti2012}. For $M_0 > 0.4$, a slight deviation between
the red and black lines emerges. For $M_0 > 0.6$, $u_{H\text{min}}$
varies significantly with $\tau$, as illustrated by the
relatively large error bars in Fig.~\ref{fig:transition}, suggesting
greater uncertainty in the precise location of the global minimum of
$H$. This region coincides with cases where the order of the phase
transition becomes unclear, as indicated by the highlighted yellow
section over the black line, according to~\citet{Benetti2012}.

We also include a green dashed line in Fig.~\ref{fig:transition}
representing a generalized virial curve $1-2u_0=M_0 \cos\theta_0$,
deduced by~\citet{Benetti2012}, corresponding to initial conditions
for which $M_x(t)\simeq M_0$ is approximately invariant and its
fluctuations are suppressed. Simulations initialized far from the
virial curve tend to experience stronger modulated fluctuations, and
some particles may undergo parametric resonances, pushing the system
further from equilibrium~\cite{Benetti2012}. As seen in
Fig.~\ref{fig:transition}, for $M_0 \lesssim 0.35$, the virial curve
closely aligns with both the curve of minimum entropy $H_{\min}$ and
the critical energy $u^\ast$, suggesting that in this regime,
different stability measures converge to a similar characteristic
energy scale. However, for larger values of $M_0>0.5$, this agreement
disappears as the virial curve is significantly separated from both
the black and red lines. This suggests that while virialization may
help suppress macroscopic fluctuations, it is not universally tied to
the fundamental structure of QSSs.

In general, the gray area below the red line is mostly characterized
by magnetized ($M_\text{QSS}\neq0$) QSSs, where the magnetization
fluctuations exhibit more ordered dynamics (lower $H$) with fewer
correlated structures (lower $C$) as $u_0$ increases. In contrast, in
the white area above this line, the QSS is demagnetized
($M_\text{QSS}=0$) and exhibits wide-amplitude magnetization
fluctuations. As $u_0$ increases, disorder grows (higher $H$), while
the complexity approaches its maximum ($C\approx C_\text{max}$),
indicating the presence of maximally correlated structures.

\section{Conclusions}
The Hamiltonian Mean-Field (HMF) model is a simple model used to
describe systems with long-range interactions. This model presents an
out-of-equilibrium phase transition separating quasi-stationary states
(QSS) characterized by either a magnetized ($M_\text{QSS}>0$) and
inhomogeneous phase-space, from a demagnetized ($M_\text{QSS}=0$) and
relatively homogeneous phase-space. While previous studies have
primarily focused on mean magnetization, macroscopic quantities such
as system temperature, or phase-space structures, the role of
magnetization fluctuations has received less attention. Here, we
investigate whether properties of the QSS states could be inferred
solely from the magnetization fluctuations. This approach is
particularly relevant in contexts where direct particle data is scarce
or inaccessible, such as satellite observations. In such cases, field
quantities like the electric field are often measurable, whereas
particle distribution functions remain incomplete or unavailable.

In this work, we characterized the QSSs of the HMF model by analyzing
the permutation entropy $H$~\cite{bandt2002} and statistical
complexity $C$~\cite{lopezruiz1995} of the magnetization time
series. Our study examined how these information-theoretic measures
vary with respect to the initial conditions and the embedding delay
$\tau$, which controls the time scale over which ordinal patterns are
extracted for the entropy calculation.

The complexity-entropy plane analysis reveals that most data points lie
above the $C$-$H$ curve for stochastic processes with a power-law
spectrum (k-noise), indicating that the magnetization time series
retains more structure than purely stochastic signals. For small
embedding delays ($\tau \leq 10$), the system exhibits low entropy and
complexity, clustering in the lower-left region of the $C$-$H$
plane. As $\tau$ increases, below the characteristic oscillation period
of the magnetization, the entropy and complexity shift toward higher
values, and the data points spread between the theoretical complexity
maximum $C_{\text{max}}$ and the k-noise curve. At large $\tau$, or
large timescales, the increased dispersion suggests a loss of ordinal
pattern coherence due to the decorrelation of distant points in the
time series.

Magnetized QSSs ($M_{\text{QSS}} \neq 0$) are characterized by
modulated, low-amplitude fluctuations of $M_x(t)$ around
$M_{\text{QSS}}$ and the formation of core-halo phase-space
distributions. Within this regime, increasing the initial energy $u_0$
leads to more ordered (lower $H$) and fewer correlated (lower $C$)
magnetization patterns. Conversely, demagnetized QSSs are
characterized by wide-amplitude fluctuations of $M_x(t)$ around
$M_{\text{QSS}} = 0$ and the formation of counter-propagating clusters
in phase space. These fluctuations also become progressively more
disordered as $u_0$ increases, with entropy increasing along the
$C_{\max}$ curve. This indicates that the system reaches a regime
where magnetization patterns exhibit the highest possible dynamical
complexity for a given entropy, suggesting that correlations persist
despite the increase in disorder.

The energies $u_0$ where $H$ reaches its global minimum align closely
with the critical energies $u^\ast$ separating magnetized and
demagnetized QSSs, particularly for $M_0 < 0.4$, where a first-order
out-of-equilibrium phase transition has been reported for the HMF
model. For $M_0 > 0.6$, where the nature of the out-of-equilibrium
phase transition is unclear~\cite{Benetti2012}, $u_{H\text{min}}$
exhibits stronger variations with $\tau$. This increased sensitivity
to $\tau$ leads to larger uncertainty estimates in determining a
well-defined global minimum of $H$. While the virial curve provides
insight into the stability of initial conditions by identifying states
where macroscopic fluctuations are suppressed, it does not universally
correlate with the structural transitions observed in QSSs. In
particular, for $M_0 > 0.4$, the virial curve deviates significantly
from both the out-of-equilibrium phase transition energy $u^\ast$ and
the entropy-minimization energy $u_{H\text{min}}$.  This suggests
that, although virialization can influence dynamical stability, it is
not directly tied to the fundamental structural changes occurring in
QSSs as $u_0$ increases.

These findings reinforce the connection between information-theoretic
measures and the out-of-equilibrium phase transitions observed in
long-range interacting systems. Future work could extend this analysis
by incorporating a more detailed time-series analysis of the full
phase-space. While such studies have already provided valuable
insights into the microscopic dynamics of the HMF
model~\cite{Bachelard2008, Benetti2012}, further exploration of the
interplay between phase-space structures and magnetization
fluctuations through the permutation entropy could enhance our
understanding of QSS characteristics. Moreover, extending these
methods to other long-range interacting systems could provide a
broader framework for analyzing out-of-equilibrium statistical
mechanics beyond the HMF model.

\providecommand{\noopsort}[1]{}\providecommand{\singleletter}[1]{#1}%

\end{document}